\documentclass[a4paper]{article}

\usepackage{INTERSPEECH2022}
\usepackage{cite}
\usepackage{subcaption}
\newsavebox{\bigimage}
\usepackage{float}
\usepackage{enumitem}
 
\usepackage[hidelinks]{hyperref}
\usepackage{hyperref}

\title{CMGAN: Conformer-based Metric GAN for Speech Enhancement}
\name{Ruizhe Cao$^*$, Sherif Abdulatif$\,^*$, Bin Yang}
\address{
  University of Stuttgart, Institute of Signal Processing and System Theory, Stuttgart, Germany
\thanks{\hspace{-2mm}\fontsize{7.3pt}{7.3pt}\selectfont\textbf{Email: ruizhe.cao96@gmail.com, sherif.abdulatif@iss.uni-stuttgart.de}
\textbf{$^*$The first two authors equally contributed to this work.}\\
\textbf{A longer version is available in \textit{\href{https://arxiv.org/abs/2209.11112}{https://arxiv.org/abs/2209.11112}}}
}
}

\begin{document}

\maketitle
\begin{abstract}
	Recently, convolution-augmented transformer (Conformer) has achieved promising performance in automatic speech recognition (ASR) and time-domain speech enhancement (SE), as it can capture both local and global dependencies in the speech signal. In this paper, we propose a conformer-based metric generative adversarial network (CMGAN) for SE in the time-frequency (TF) domain. In the generator, we utilize two-stage conformer blocks to aggregate all magnitude and complex spectrogram information by modeling both time and frequency dependencies. The estimation of magnitude and complex spectrogram is decoupled in the decoder stage and then jointly incorporated to reconstruct the enhanced speech. In addition, a metric discriminator is employed to further improve the quality of the enhanced estimated speech by optimizing the generator with respect to a corresponding evaluation score. Quantitative analysis on Voice Bank+DEMAND dataset indicates the capability of CMGAN in outperforming various previous models with a margin, i.e., PESQ of 3.41 and SSNR of 11.10 dB. 

\end{abstract}
\noindent\textbf{Index Terms}: Speech enhancement, conformer, transformer, generative adversarial networks, metric discriminator.

\vspace{-3mm}\section{Introduction}\vspace{-1mm}
In realistic use cases, the perceived speech quality and intelligibility are directly dependent on the performance of the underlying speech enhancement (SE) systems. As such, SE frameworks are an indispensable component in modern automatic speech recognition (ASR), telecommunication systems and hearing aid devices. This is evident by the increasingly large amount of research publications continuously attempting to push the performance boundaries of current SE systems \cite{wang2018supervised, loizou2007speech}. The majority of these approaches harness the recent advances in deep learning (DL) techniques as well as the increasingly more available speech datasets \cite{valentini2016investigating, article, barker2015third, dubey2022icassp}.

The aforementioned body of works for SE techniques can be roughly categorized into two prominent families of approaches. Chronologically, enhancing the speech time-frequency (TF) representations constitute the classical SE paradigm which encompasses the majority of model-based as well as more recent DL approaches \cite{wang2018supervised, fu2019metricgan, yin2020phasen, yu2022dualbranch}. More recently, a new set of approaches were introduced to enhance raw waveforms directly without any transformational overheads \cite{pascual2017segan, macartney2018improved, 9413740,defossez2020real, kim21h_interspeech}. Each paradigm presents unique advantages and drawbacks.

The time-domain paradigm is based on generative models trained to directly estimate fragments of the clean waveform from the noisy counterparts. In this case, the whole information is preserved in the waveform without any transformation or reconstruction requirements \cite{macartney2018improved, 9413740}. However, the lack of direct frequency representation hinders these frameworks from capturing speech phonetics in the frequency domain. This limitation is usually reflected as artifacts in the reconstructed speech. Another drawback of this paradigm is the ample input space associated with the raw waveforms, which often necessitates the utilization of deep computationally complex frameworks \cite{pascual2017segan, defossez2020real}.

In the TF-domain, conventional model-based or DL techniques would utilize the magnitude component while ignoring the phase. This is accounted to the random structure in the phase component, which imposes challenges to the utilized architectures \cite{aegan,invest}. To circumvent the challenging phase, recent approaches follow the strategy of enhancing the complex spectrogram, which implicitly enhances both magnitude and phase. However, the compensation effect between the magnitude and phase often leads to an inaccurate magnitude estimation \cite{wang2021compensation}. Moreover, the prediction of complex spectrogram by either using complex ratio masking \cite{complex_ratio_masking} or direct mapping often leads to an unbounded estimation, which degrades the enhancement performance. More recent studies proposed enhancing the magnitude followed by complex spectrogram refinement, which can alleviate these problems effectively \cite{yu2022dualbranch, li2022glance}. Furthermore, the commonly used objective function in SE is simply the $L^p-$norm distance between the estimated and the target spectrograms. Nevertheless, a lower distance does not always lead to higher speech quality. MetricGAN is proposed to overcome this issue by optimizing the generator with respect to the evaluation metric score, which is learned by a discriminator \cite{fu2019metricgan}. In addition, many approaches utilize transformers \cite{vaswani2017attention} to capture the long-distance dependency in the waveform or the spectrogram \cite{yu2022dualbranch, 9413740, dang2021dpt}. Recently, conformers have been introduced as an alternative to transformers in ASR and speech separation tasks due to their capacity of capturing both local context and global context \cite{gulati2020conformer,chen2021continuous}. Accordingly, they were also employed for time-domain SE \cite{kim21h_interspeech}. To the best of our knowledge, conformers are not yet explicitly investigated for TF-domain SE.

\begin{figure*}[t]
	\includegraphics[width=\textwidth]{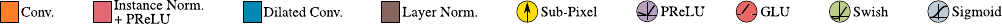}
	\vspace{0mm}
	\centering
	
	\sbox{\bigimage}{%
		\begin{subfigure}[b]{.485\textwidth}
			\centering
			\includegraphics[width=\textwidth]{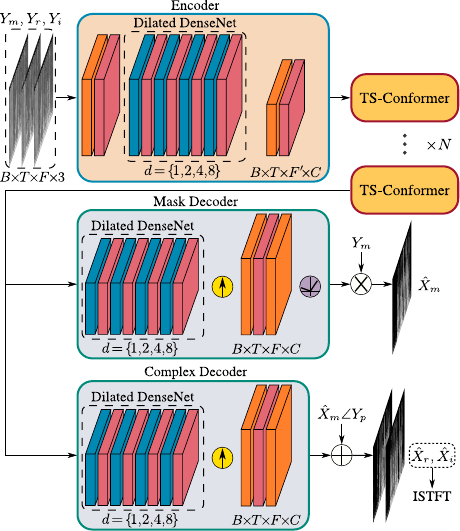}
			\caption{Encoder-decoders generator architecture.}
			\label{fig:generator}
			\vspace{0pt}
		\end{subfigure}%
	}
	
	\usebox{\bigimage}\hfill 
	\begin{minipage}[b][\ht\bigimage][s]{.485\textwidth}
		\begin{subfigure}{\textwidth}
			\centering
			\includegraphics[height=6.476cm,width=\textwidth]{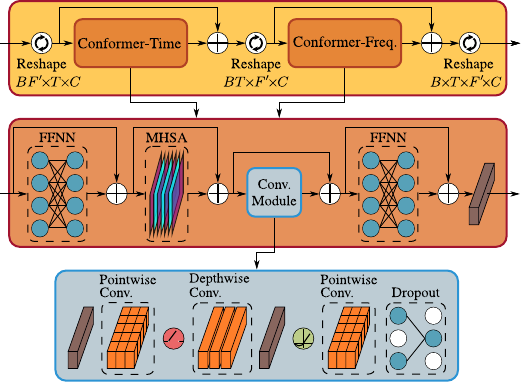}
			\caption{Two-stage conformer (TS-Conformer).}
			\label{fig:TS-Conformer}
		\end{subfigure}%
		\vfill
		\begin{subfigure}{\textwidth}
			\centering
			\includegraphics[height=2.418cm,width=\textwidth]{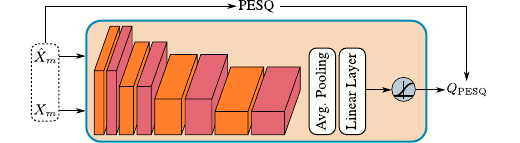}
			\caption{Metric discriminator.}
			\label{fig:discriminator}
		\end{subfigure}
		
		\vspace{0pt}
	\end{minipage}
	\vspace{-2mm}
	\caption{An overview of the proposed CMGAN architecture. }
	\label{Overview}
	\vspace{-3mm}
\end{figure*}

Inspired by the stated problems and previous works, we propose a conformer-based MetricGAN (CMGAN) for monaural speech enhancement. The CMGAN consists of a generator and a metric discriminator. The generator is based on two-stage conformer blocks in the TF-domain, while the discriminator is responsible for estimating a black-box non-differentiable metric. The concatenated magnitude and complex components are passed to the generator, which comprises an encoder with two-stage conformer blocks, a mask decoder and a complex decoder. The encoder aims to learn a compact feature representation of the input. The mask decoder estimates the mask for the input magnitude and the complex decoder estimates the compensation for real and imaginary components. In order to reduce the significant computational complexity of the conformer, we adopt the dual-path transformers \cite{9413740, dang2021dpt,chen2020is} into a two-stage conformer block, which can capture the dependencies along the time dimension and the frequency dimension sequentially. In a nutshell, the contributions of this work are three-fold:
\begin{itemize}[label=$\bullet$,wide = 0pt]
	\vspace{-1mm}
	\setlength{\itemindent}{0.5em}
	\setlength{\itemsep}{1.5pt}
	\setlength{\parskip}{1pt}
    \item We investigate the performance of the two-stage conformer blocks and its capability of capturing time and frequency dependencies with a relatively low computational complexity.
    \item We adopt a metric discriminator to our network, which helps to improve the corresponding evaluation metric without adversely affecting other metrics.
    \item The proposed approach outperforms other former approaches on the Voice Bank+DEMAND dataset \cite{valentini2016investigating}. A comprehensive ablation study verifies the effectiveness of our design choices.
\end{itemize}

\section{Methodology}
\subsection{Generator architecture}
An overview of the generator architecture is shown in Fig.~\ref{fig:generator}. For a noisy speech waveform $y\in \mathbb{R}^{L\times 1}$, a short-time Fourier transform (STFT) first converts the waveform into a complex spectrogram $Y_{o}\in \mathbb{R}^{T \times F\times 2}$, where $T$ and $F$ denote the time and frequency dimensions, respectively. Then the compressed spectrogram $Y$ is obtained by the power-law compression:
\begin{equation}
    Y=|Y_o|^ce^{jY_p} = Y_me^{jY_p}=Y_r+jY_i
\end{equation}
where $Y_m$, $Y_p$, $Y_r$ and $Y_i$ denote the magnitude, phase, real and imaginary components of the compressed spectrogram, respectively. $c$ is the compression exponent which ranges from 0 to 1, here we follow Braun \textit{et al.} \cite{braun2020consolidated} to set $c=0.3$. The power-law compression of the magnitude equalizes the importance of quieter sounds relative to loud ones, which is closer to human perception of sound \cite{8392411, wilson2018exploring}. The real and imaginary parts $Y_r$ and $Y_i$ are then concatenated with the magnitude $Y_m$ as an input to the generator.

\vspace{-1.5mm}
\subsubsection{Encoder}
Given the input feature $Y_{in} \in \mathbb{R}^{B\times T\times F\times 3}$, where $B$ denotes batch size, the encoder consists of two convolution blocks with a dilated DenseNet \cite{pandey2020densely} in between. Each convolution block comprises a convolution layer, an instance normalization \cite{ulyanov2016instance} and a PReLU activation \cite{he2015delving}. The first convolution block is used to extend the three input features to an intermediate feature map with $C$ channels. The dilated DenseNet contains four convolution blocks with dense connections, the dilation factors of each block are set to \{1, 2, 4, 8\}. The dense connections can aggregate all previous feature maps to extract different feature levels. As for the dilated convolutions, they serve to increase the receptive field effectively while preserving the kernels and layers count. The last convolution block is responsible for halving the frequency dimension to $F'$ to reduce the complexity.

\vspace{-1.5mm}
\subsubsection{Two-stage conformer block}
Conformers \cite{gulati2020conformer,chen2021continuous} achieved great success in speech recognition and separation as they combine the advantages of both transformers and convolutional neural networks (CNNs). Transformers can capture long-distance dependencies, while CNNs exploit local features effectively. Here we employ two conformer blocks sequentially to capture the time dependency in the first stage and the frequency dependency in the second stage. As shown in the Fig.~\ref{fig:TS-Conformer}, given a feature map $D\in \mathbb{R}^ {B \times T \times F' \times C}$, the input feature map $D$ is first reshaped to $D^{T} \in \mathbb{R}^{BF'\times T \times C}$ to capture the time dependency in the first conformer block. Then the output $D^{T}_o$ is element-wise added with the input $D^T$ (residual connection) and reshaped to a new feature map $D^{F}\in \mathbb{R}^{BT\times F' \times C}$. The second conformer thus captures the frequency dependency. After the residual connection, the final output $D_o$ is reshaped back to the input size.

Similar to \cite{gulati2020conformer}, each conformer block utilizes two half-step feed-forward neural networks (FFNNs). Between the two FFNNs, a multi-head self-attention (MHSA) with four heads is employed, followed by a convolution module. The convolution module depicted in Fig.~\ref{fig:TS-Conformer} starts with a layer normalization, a point-wise convolution layer and a gated linear unit (GLU) activation to diminish the vanishing gradient problem. The output of the GLU is then passed to a 1D-depthwise convolution layer with a swish activation function, then another point-wise convolution layer. Finally, a dropout layer is used to regularize the network. Also, a residual path connects the input to the output.

\subsubsection{Decoder}
The decoder extracts the output from $N$ two-stage conformer blocks in a decoupled way, which includes two paths: the mask decoder and the complex decoder. The mask decoder aims to predict a mask that will be element-wise multiplied by the input magnitude. On the other hand, the complex decoder directly predicts the real and imaginary parts. Both mask and complex decoders consist of a dilated DenseNet, similar to the one in the encoder. The subpixel convolution layer is utilized in both paths to upsample the frequency dimension back to $F$. For the mask decoder, a convolution block is used to squeeze the channel number to 1, followed by another convolution layer with PReLU activation to predict the final mask. Note that the PReLU activation learns different slopes for each frequency band. Post-training evaluation indicates that the slopes reflect negative values, i.e., the output mask is always projected in the positive 1\textsuperscript{st} and 2\textsuperscript{nd} quadrants. For the complex decoder, the architecture is identical to the mask decoder, except no activation function is applied for the complex output since it is unbounded.

Same as in \cite{yu2022dualbranch, li2022glance}, the masked magnitude $\hat{X}_m$ is first combined with the noisy phase $Y_p$ to obtain the magnitude-enhanced complex spectrogram, then it is element-wise summed with the output of the complex decoder $(\hat{X}'_r, \hat{X}'_i)$ to obtain the final complex spectrogram: \vspace{-1.0mm}
\begin{equation}
    \hat{X}_r=\hat{X}_mcos(Y_p)+\hat{X}'_r \hspace{8mm}
    \hat{X}_i=\hat{X}_msin(Y_p)+\hat{X}'_i
\end{equation}
The power-law compression is then inverted on the final complex spectrogram $(\hat{X}_r, \hat{X}_i)$ and an inverse short-time Fourier transform (ISTFT) is applied to get the time-domain signal $\hat{x}$.
\vspace{-1mm}
\subsection{Metric discriminator}
In SE, the objective functions are often not directly correlated to the evaluation metrics. Consequently, even if the objective loss is optimized, the evaluation score is still not satisfied. Furthermore, some evaluation metrics like perceptual evaluation of speech quality (PESQ) \cite{941023} and short-time objective intelligibility (STOI) \cite{taal2010short} can not be used as loss functions because they are non-differentiable. Hence, the discriminator in CMGAN aims to mimic the metric score and use it as a part of the loss function. Here we follow the MetricGAN to use the PESQ score as a label \cite{fu2019metricgan}. As shown in Fig.~\ref{fig:discriminator}, the discriminator consists of 4 convolution blocks. Each block starts with a convolution layer, followed by instance normalization and a PReLU activation. After the convolution blocks, a global average pooling is followed by two feed-forward layers and a sigmoid activation. The discriminator is then trained to estimate the maximum normalized PESQ score ($=1$) by taking both inputs as clean magnitudes. Additionally, the discriminator is trained to estimate the enhanced PESQ score by taking both clean and enhanced spectrum as an input together with their corresponding PESQ label. On the other hand, the generator is trained to render an enhanced speech resembling the clean speech, thus approaching a PESQ label of 1.

\vspace{-1mm}
\subsection{Loss function}
Inspired by Braun \textit{et al.} \cite{braun2020consolidated}, we use the linear combination of magnitude loss $\mathcal{L}_{\small\textrm{Mag.}}$ and complex loss $\mathcal{L}_{\small\textrm{RI}}$ in the TF-domain:
\begin{equation}
\begin{aligned}
    &\mathcal{L}_{\small\textrm{TF}} =\alpha\, \mathcal{L}_{\small\textrm{Mag.}}+ (1-\alpha)\,\mathcal{L}_{\small\textrm{RI}} \\
    &\mathcal{L}_{\small\textrm{Mag.}} =\mathbb{E}_{X_m,\hat{X}_m} \big[\| X_m - \hat{X}_m  \|^2 \big]  \\
    &\mathcal{L}_{\small\textrm{RI}} =\mathbb{E}_{X_r,\hat{X}_r} \big[\| X_r - \hat{X}_r \|^2\big] + \mathbb{E}_{X_i,\hat{X}_i} \big[\| X_i - \hat{X}_i \|^2\big]
\end{aligned}
\label{TF loss}
\end{equation}
where $\alpha$ is a chosen weight, based on grid search, $\alpha=0.7$ leads to the best performance.
Similar to least-square GANs \cite{lsgans}, the adversarial training is following a min-min optimization task over the discriminator loss $\mathcal{L}_{\small\textrm{D}}$ and the corresponding generator loss $\mathcal{L}_{\small\textrm{GAN}}$ expressed as follows:
\begin{equation}
\begin{aligned}
    &\mathcal{L}_{\small\textrm{GAN}} =\mathbb{E}_{X_m,\hat{X}_m} \big[\| D(X_m,\hat{X}_m) - 1 \|^2\big]\\
    &
    \begin{split}
    	\mathcal{L}_{\small\textrm{D}} &= \mathbb{E}_{X_m} \big[\| D(X_m, X_m) - 1\|^2\big] \\&+ \mathbb{E}_{X_m,\hat{X}_m}\big[\,\| D(X_m,\hat{X}_m)- Q_{\small\textrm{PESQ}} \|^2\big] 
    \end{split}
\end{aligned}
\vspace{-0.25mm}
\label{GAN loss}
\end{equation}
where $D$ refers to the discriminator and $Q_{\small\textrm{PESQ}}$ refers to the normalized PESQ score. Here we normalize the PESQ score to the range [0,1]. Moreover, an additional penalization in the resultant waveform $\mathcal{L}_{\small\textrm{Time}}$ is proven to improve the restored speech quality \cite{invest}:
\begin{equation}
    \mathcal{L}_{\small\textrm{Time}}= \mathbb{E}_{x,\hat{x}} \big[\| x-\hat{x} \|_1\big]
    \vspace{-0.25mm}
    \label{time loss}
\end{equation}
where $\hat{x}$ is the enhanced waveform and $x$ is the clean target waveform. The final generator loss is formulated as follows:
\begin{equation}
    \mathcal{L}_{\small\textrm{G}}=\gamma_1 \,\mathcal{L}_{\small\textrm{TF}} + \gamma_2 \,\mathcal{L}_{\small\textrm{GAN}} + \gamma_3 \,\mathcal{L}_{\small\textrm{Time}}
    \vspace{-0.25mm}
\end{equation}
where $\gamma_1, \gamma_2$ and $\gamma_3$ are the weights of the corresponding losses and they are chosen to reflect equal importance.
\vspace{-1mm}
\section{Experiments}
\vspace{-1mm}
\subsection{Datasets}
\vspace{-1mm}
For quantitative analysis, we investigate our proposed approach on the commonly used publicly available Voice Bank+DEMAND dataset \cite{valentini2016investigating}. The clean tracks are selected from the Voice Bank corpus \cite{6709856} which includes 11,572 utterances from 28 speakers in the training set and 824 utterances from 2 unseen speakers in the test set. In the training set, the clean utterances are mixed with background noise (8 noise types from DEMAND database \cite{thiemann2013diverse} and 2 artificial noise types) at SNRs of 0 dB, 5 dB, 10 dB and 15 dB. In the test set, the clean utterances are mixed with 5 unseen noise types from the DEMAND database at SNRs of 2.5 dB, 7.5dB, 12.5 dB and 17.5 dB. All utterances are resampled to 16 kHz in our experiments.

\begin{table*}[t]
\centering
\caption{Performance comparison on Voice Bank+DEMAND dataset \cite{valentini2016investigating}. “-” denotes the result is not provided in the original paper.}
\begin{tabular}{lllcccccccc}
\toprule
Method  & Year & Input & Model Size (M) &  PESQ & CSIG & CBAK & COVL & SSNR & STOI  \\
\midrule
Noisy  & \hspace{2.5mm}- & \hspace{2.5mm}- & -&  1.97 & 3.35 & 2.44 & 2.63 & 1.68 & 0.91  \\
\midrule
SEGAN \cite{pascual2017segan} & 2017 & Time & 97.47& 2.16 & 3.48 & 2.94 & 2.80 & 7.73 & 0.92  \\
MetricGAN \cite{fu2019metricgan} & 2019 & Magnitude & - & 2.86 & 3.99 & 3.18 & 3.42 & - & - \\
PHASEN \cite{yin2020phasen} & 2020 & Magnitude+Phase & - & 2.99 & 4.21 & 3.55 & 3.62 & 10.08 & -  \\
TSTNN \cite{9413740} & 2021 & Time & 0.92& 2.96 & 4.10 & 3.77 & 3.52 & 9.70 & 0.95 \\
DEMUCS \cite{defossez2020real} & 2021 & Time & 128& 3.07 & 4.31 & 3.40 & 3.63 & - & 0.95  \\
PFPL \cite{hsieh2020improving} & 2021 & Complex & - & 3.15 & 4.18 & 3.60 & 3.67 & - & 0.95  \\
MetricGAN+ \cite{fu2021metricgan+} & 2021 & Magnitude & - & 3.15 & 4.14 & 3.16 & 3.64 & - & -  \\
SE-Conformer \cite{kim21h_interspeech} & 2021 & Time & -& 3.13 & 4.45 & 3.55 & 3.82 & - & 0.95  \\
DB-AIAT \cite{yu2022dualbranch} & 2021 & Complex+Magnitude & 2.81& 3.31 & 4.61 & 3.75 & 3.96 & 10.79 & 0.96  \\
DPT-FSNet \cite{dang2021dpt} & 2021 & Complex & 0.91& 3.33 & 4.58 & 3.72 & 4.00 & - &  0.96  \\
\textbf{CMGAN} & 2022 & Complex+Magnitude & 1.83& \textbf{3.41} & \textbf{4.63} & \textbf{3.94} & \textbf{4.12} & \textbf{11.10} & \textbf{0.96}  \\
\bottomrule
\end{tabular}
\label{result comparison demand}
\vspace{-3mm}
\end{table*}

\begin{table}[b]
	\vspace{-3mm}
\centering
\small
\caption{Results of the ablation study}
\setlength{\tabcolsep}{0.5mm}
\begin{tabular}{lcccccc}
\toprule  
Method  & PESQ & CSIG & CBAK & COVL & SSNR & STOI \\
\midrule
CMGAN & 3.41 & \textbf{4.63} & \textbf{3.94} & \textbf{4.12} & \textbf{11.10} & \textbf{0.96} \\
Magnitude-only & 3.23 & 4.60 & 3.76 & 4.00 & 9.82 & 0.95 \\
Complex-only & 3.35 & 4.56 & 3.79 & 4.05 & 9.19 & 0.96 \\
w/o Time loss & \textbf{3.45} & 4.56 & 3.86 & 4.11 & 9.71 & 0.96 \\
w/o Disc. & 3.24 & 4.46 & 3.82 & 3.93 & 10.56 & 0.96 \\
Patch Disc. & 3.28 & 4.48 & 3.85 & 3.96 & 10.75 & 0.96 \\
Parallel-Conformer & 3.35 & 4.54 & 3.87 & 4.03 & 10.63 & 0.96 \\
Freq. $\rightarrow$ Time & 3.39 & 4.56 & 3.91 & 4.07 & 10.84 & 0.96 \\
\bottomrule 
\end{tabular}
\label{ablation}
\end{table}
\vspace{-1mm}
\subsection{Experimental setup}
\vspace{-1mm}
The utterances in the training set are sliced into 2 seconds, while in the test set, no slicing is utilized and the length is kept variable. A Hamming window with 25 ms window length (400-point FFT) and hop size of 6.25 ms (75\% overlap) is employed. Thus, the resultant spectrogram will have 200 frequency bins $F$, while the time dimension $T$ depends on the variable track duration.
The number of two-stage conformer blocks $N$, the batch size $B$ and the channel number $C$ in the generator are set to 4, 4 and 64, respectively. The channel numbers in the metric discriminator are set to \{16, 32, 64, 128\}. In the training stage, AdamW optimizer \cite{loshchilov2017decoupled} is used for both the generator and the discriminator to train for 50 epochs. Audio samples are made available online\footnote{\textit{\href{https://github.com/ruizhecao96/CMGAN/}{https://github.com/ruizhecao96/CMGAN/}}}and all implementations of CMGAN will be made publicly available upon the publication of this work.
\vspace{-0.5mm}
\section{Results and discussion} 
We choose a set of commonly used metrics to evaluate the enhanced speech quality, i.e., PESQ with a score range from -0.5 to 4.5, segmental signal-to-noise ratio (SSNR) and mean opinion score (MOS) \cite{hu2007evaluation} based metrics: MOS prediction of the signal distortion (CSIG), MOS prediction of the intrusiveness of background noise (CBAK) and MOS prediction of the overall effect (COVL), all of them are within a score range of 1 to 5. Additionally, we utilize STOI with a score range from 0 to 1 to judge speech intelligibility. Higher values indicate better performance for all metrics. 

\subsection{Results analysis}
Our proposed CMGAN is objectively compared with other state-of-the-art (SOTA) baselines as shown in Table~\ref{result comparison demand}. For the time-domain methods, we included the standard SEGAN \cite{pascual2017segan} and three recent methods: TSTNN \cite{9413740}, DEMUCS \cite{defossez2020real} and SE-Conformer \cite{kim21h_interspeech}. For the TF-domain methods, we evaluate six recent SOTA methods, i.e., MetricGAN \cite{fu2019metricgan}, PHASEN \cite{yin2020phasen}, PFPL \cite{hsieh2020improving}, MetricGAN+ \cite{fu2021metricgan+}, DB-AIAT \cite{yu2022dualbranch} and DPT-FSNet \cite{dang2021dpt}. It can be observed that most of the TF-domain methods outperform the time-domain counterparts over all utilized metrics. Moreover, our proposed TF conformer-based approach shows a major improvement over the time-domain SE-Conformer. Compared to frameworks involving metric discriminators (MetricGAN+), we have 0.26, 0.49, 0.78 and 0.48 improvements on the PESQ, CSIG, CBAK and COVL scores, respectively. Finally, our framework also outperforms recent improved transformer-based methods, such as DB-AIAT and DPT-FSNet in all of the evaluation scores with a relatively low model size of only 1.83~M parameters.

\subsection{Ablation study}

To verify our design choices, an ablation study is conducted as shown in Table~\ref{ablation}. We first investigate the influence of different inputs. Magnitude-only denotes only magnitude is used as the input and the enhanced magnitude is then combined with the noisy phase for ISTFT operation. The network architecture remains the same, except the complex decoder is removed. Correspondingly, Complex-only denotes only complex spectrogram is used as an input and the mask decoder is removed. Comparison between them shows that lacking the phase enhancement decreases the PESQ score by 0.18, while using pure complex spectrogram reduces the SSNR score by 1.91 dB. This result indicates that although the complex spectrogram contains magnitude information, it is challenging for the utilized framework to enhance the magnitude implicitly.

On the other hand, the result shows that the absence of time loss (w/o Time loss) further improves the PESQ score to 3.45, while the SSNR is slightly lower than the original CMGAN. This indicates the effectiveness of the time loss in balancing the performance for both PESQ and SSNR scores. We conducted two tests to demonstrate the discriminator choice: removing the discriminator (w/o Disc.) and replacing the metric discriminator with a patch discriminator, which is commonly used in image generation tasks \cite{pix2pix}. It can be realized that removing the discriminator negatively impacted all the given scores. Furthermore, adding a patch discriminator only showed a marginal improvement, which reflects that the generator is fully capable of enhancing the tracks without the aid of a normal patch discriminator. However, a metric discriminator to directly improve the evaluation scores is proven to be beneficial.

Finally, we investigate the influence of the two-stage conformer outline. Given an input feature map, the two-stage conformer will separately focus on the time and frequency dimensions. To this end, two different configurations can be proposed, either sequential or parallel. Accordingly, we compare our sequential CMGAN to a parallel connection counterpart without any further modification (Parallel-Conformer). The results illustrate that the parallel approach is behind the proposed sequential, i.e., the PESQ and SSNR scores are reduced by 0.06 and 0.47 dB, respectively. Also, we flipped the order of the sequential conformer blocks (Freq. $\rightarrow$ Time) and we can conclude that the scores are similar with a minor improvement in favor of the standard CMGAN (Time $\rightarrow$ Freq.).
\vspace{-1.5mm}
\section{Conclusions}
In this work, we introduce CMGAN for speech enhancement operating on both magnitude and complex spectrogram components. Our approach combines recent conformers that can capture long-term dependencies as well as local features in both time and frequency dimensions, together with a metric discriminator that resolves metric mismatch by directly enhancing non-differentiable evaluation scores. Experimental results show that the proposed method outperforms the current SOTA on the Voice Bank+DEMAND dataset with relatively few parameters (1.83~M). Additionally, an ablation study is conducted to verify the fragmented benefits of each utilized component and loss in the proposed CMGAN framework. In the future, a subjective evaluation study should be conducted with real listening tests. Furthermore, our study should be expanded to involve other speech enhancement tasks, such as dereverberation and audio superresolution.

\clearpage
\bibliographystyle{IEEEtran}

\end{document}